\documentclass[12pt,a4paper]{article}
\usepackage{todonotes}
\usepackage{authblk}
\usepackage{amsthm}
\usepackage[english]{babel}
\usepackage{graphicx}
\usepackage[latin1]{inputenc}
\usepackage{verbatim}
\usepackage{amsfonts}
\usepackage{amsmath}
\usepackage{amssymb}
\usepackage[T1]{fontenc}
\usepackage{color}
\usepackage{epsfig}
\usepackage{natbib}
\usepackage{url}
\usepackage[bookmarksnumbered=true, colorlinks=true, citecolor=blue]{hyperref}
\date{}
\newtheorem{ten}{Tenet}

\title{\bf {On the Prospects of a de Broglie-Bohm-Barbour-Bertotti Theory}}
\author[a]{Antonio Vassallo
\thanks{\href{mailto:antonio.vassallo1977@gmail.com}{antonio.vassallo1977@gmail.com}}}
\author[a,b]{Pedro Naranjo
\thanks{\href{mailto:pnpfisica@gmail.com}{pnpfisica@gmail.com}}}
\affil[a]{Warsaw University of Technology, Faculty of Administration and Social Sciences, Plac Politechniki 1, 00-661 Warsaw}
\affil[b]{University of Warsaw, Faculty of Philosophy, Krakowskie Przedmie\'scie 3, 00-047 Warsaw}

\begin{document}
\maketitle
%\vspace{-3mm}
\begin{center}
Accepted for publication in A. Oldofredi (Ed.), \emph{Guiding Waves in Quantum Mechanics: 100 Years of de Broglie-Bohm Pilot-Wave Theory}, Oxford University Press.
\end{center}
\vspace{3mm}
\begin{abstract}
Pure shape dynamics (PSD) is a novel implementation of the relational framework originally proposed by Julian Barbour and Bruno Bertotti. PSD represents a Leibnizian/Machian approach to physics in that it completely describes the dynamical evolution of a physical system without resorting to any structure external to the system itself. The chapter discusses how PSD effectively describes a de Broglie-Bohm $N$-body system, and the conceptual benefits of such a relational description. The analysis will highlight the new directions in the quest for an understanding of the nature of the wave function that are opened up by a modern relationalist elaboration on de Broglie's and Bohm's original insights.
\end{abstract}

%This is achieved by casting the system's dynamics in terms of the geometric properties of an unparametrized curve defined over the appropriate relational configuration space (called shape space). This dynamical description marks the essential difference between PSD and standard shape dynamics (SD): Whereas SD  involves a notion of parametrized curve, which translates into an extra degree of freedom external to the system, PSD dispenses with the need for a parametrization altogether. 
\newpage
\tableofcontents

\section{Introduction}
The modern landscape of fundamental physics is dominated by quantum theories. Non-relativistic quantum mechanics is used to describe low-energy systems, while quantum field theory is the standard framework for high-energy physics. Quantum theories are so empirically successful that, nowadays, there is a wide consensus about the fact that the fundamental description of the gravitational field must be quantum as well. 

However, all quantum theories---whether relativistic or not---have to face a plethora of conceptual problems, irrespective of their predictive success. In particular, there are two questions that the quantum formalism used by physicists is unable to address: First, how can the \emph{stable} and \emph{determinate} features of the world, such as macroscopic material structures, be accounted for by a dynamics that just deals with the linear and unitary evolution of the wave function? Second, what is the nature of the wave function itself? Is it a concrete object, like a field, or just a compact way to describe how quantum systems behave (including our expectations about measurement outcomes)?

The consequence of these issues is that the picture of the physical world entailed by quantum theories is somewhat obscure, which is a severe drawback if we want to take these theories at face value in a scientific realist spirit. Quoting a recent overview of the foundations of quantum mechanics:

\begin{quote}
[M]any physicists have come to believe that fundamental physics is in a state of stagnation, with little meaningful progress having been made in the last few decades [reference omitted]. It seems entirely possible that this has arisen because physicists never properly got to grips with what quantum mechanics tells us about the world, and therefore all subsequent physics has been based on an improper understanding of the earlier theory, leading us into a dead end \citep[][p.2]{adlam}.
\end{quote}

Usually, this ``improper understanding'' is taken to imply the need for an \emph{interpretation}---which, simply speaking, means supplying the quantum formalism with a clear metaphysics (see, \citealp{585}, for a recent textbook on this topic). In this vein, the work of Louis de Broglie \citep{338, 220} and David Bohm \citep{187a,187b} contributed a great deal to overcome the conceptual issues that plague the ``orthodox'' take on quantum mechanics. In doing so, de Broglie and Bohm shed a new light on quantum physics, especially as far as the problem of bridging the conceptual gap between the Schr\"odinger-like dynamics and the existence of stable and determinate features of the world is concerned. The way the de Broglie-Bohm (dBB) approach to quantum physics achieves that is nowadays very well known and understood (see, e.g., Passon's and Bricmont's chapters in this volume). In a nutshell, the key insight is that quantum physics is not just about the wave function: There are extra degrees of freedom that account for the physical development of material structures in spacetime---e.g., the spatial degrees of freedom of material particles. In this sense, the dBB approach comes to grips with the first question mentioned above. However, the question concerning the nature of the wave function remains open in this context. This is because the dBB approach just introduces a ``guidance principle'' according to which the wave function \emph{qua} mathematical object determines the equations of motion for the material degrees of freedom, but it leaves the question open as to what kind of physical mechanism this ``pilot wave'' picture refers to.

Indeed, two different broad views on the wave function can be recognized in the dBB context. Some people think that this piece of the formalism is best understood as referring to some kind of field-like entity that physically determines in a straightforward sense the evolution of the material degrees of freedom (\citealp{187a,187b} is the most famous example; more recent proposals include \citealp{376,687}), while others argue that the wave function is just a way to codify the physical information representing the evolution of the material degrees of freedom \citep{196,428}. In this chapter, we will argue that, in order to better understand what the nature of the wave function is in a dBB setting, a further step away from ``orthodox'' physics is needed. This different perspective on physics is represented by the modern Leibnizian/Machian take on dynamics.

As the designation suggests, this approach to dynamics is rooted in age-old relational and empiricist ideas, which have been revived and adapted to a modern physical framework starting with the seminal work of Julian Barbour and Bruno Bertotti \citep{bar, 84, 83}. The main realization behind the framework is that an empirically adequate ontology of the physical world should avoid structures whose variation makes no physical difference or which lack explanatory power to account for said differences. The immediate consequence of this metaphysical tenet is that the dynamical description of a closed system should be given in \emph{intrinsic} terms, namely, in terms of relative changes between the degrees of freedom of the system only. This is because Leibnizian/Machian relationalism avoids reference structures external to the system, which are there only to account for some sort of absolute change. The obvious examples of such structures are Newtonian absolute space and time. 

This abhorrence for external reference structures further implies that each and every part of a closed system acquires a physical characterization by virtue of standing in some type of physical relations to the rest of the system. A straightforward example of this is the Machian take on inertia: According to Mach, the origin of inertia is not local, that is, inertia is not inherent in the object of which this property is predicated. Rather, inertia accounts for how a material body is dynamically related to the rest of the universe. Note how this ``holistic'' perspective nicely fits the dBB setting, where the fundamental dynamical description concerns the universal material system---with a description in terms of subsystems that is attained only as an appropriate approximation.

In the present context, a Leibnizian/Machian approach can help in figuring out to what extent the wave function can be seen as external to a quantum system (and, thus, a ``foreign'' structure to be dispensed with in favor of a fully intrinsic description) or, in some way, part of the quantum system itself (meaning there is something physically tangible about it). The chapter will attempt at answering this question by introducing the Leibnizian/Machian approach to physics (\S~\ref{cm}) and discussing some possible quantum implementations of the framework that exploit the dBB approach (\S~\ref{qm}). In particular, the latest version of a de Broglie-Bohm-Barbour-Bertotti model will be discussed in detail (\S~\ref{dBBtechnical}), highlighting some interesting metaphysical morals regarding the nature of the wave function (\S~\ref{dBBphilosophical}). Finally, some intriguing future lines of research will be presented in \S~\ref{concl}.

\section{The Modern Leibnizian/Machian Approach to Physics}\label{Mach}

%In this section we shall provide a concise overview of the main recent developments that have occurred in the relational physics programme, with special focus on current efforts to extend the framework to the quantum realm. For completeness, we will rather briefly cover the classical case, which is well analysed elsewhere (see \citep{vasnarkos1} and references therein).

\subsection{Classical Mechanics}\label{cm}

Modern relational dynamics may be summarised in the following two tenets \citep{vasnarkos1}:

\begin{ten}[\textbf{Spatial relationalism}]\label{MachP2}
The only physically objective spatial information of a physical system is encoded in its shape, intended as its dimensionless and scale-invariant relational configuration.
\end{ten}

\begin{ten}[\textbf{Temporal relationalism}]\label{MachP3} 
Temporal structures, such as chronological ordering, duration, and temporal flow, must be defined only in terms of changes in the relational configurations of physical systems. 
\end{ten} 

The motivation for tenets \ref{MachP2} and \ref{MachP3} stems from the relationalist desire to eliminate redundant structures from physical theories, which typically arises because of the presence of various symmetries in the representation of a given physical system. This distaste for redundant structures may be dressed with heavy empiricist overtones by claiming that \emph{all structures whose variation amounts to no empirically observable difference should be shunned from physics}.

%Historically, Barbour and Bertotti consider the \emph{relative configuration space}, arrived at after quotienting by translations $\mathsf{T}$ and rotations $\mathsf{R}$, while keeping size. Since the transition from Barbour-Bertotti models to modern Shape Dynamics (SD) has been accounted for, in order to keep the discussion compact, we shall frame our discussion in terms of the latter, 

%\footnote{It is beyond the scope of this short survey to go into the technical details of SD. The interested reader is advised to look at the nice introduction in \cite{720} as well as the comprehensive, yet pedagogical book by .}

% SD is a theory of gravity (classical and relativistic)  SD is a theory of conformal 3-geometries, as opposed to standard Einstein's gravity, which is a theory of Riemannian 4-geometries. In other words, the dynamical variables of the theory are only the parts of a Riemannian $3$-metric that determine angles. 

% \footnote{Since we are here providing a brief overview of modern relationalism, we refrain from commenting further on this point. See \cite{724,723,725}, for a discussion on the possible consequences of comparativism towards physical magnitudes.} 

The work, pioneered by Barbour and Bertotti, on a theory that fully complies with the relational tenets \ref{MachP2} and \ref{MachP3} eventually led to Shape Dynamics (SD; see, e.g., \citealp{514}). According to this theory, any closed (or, universal) material system should be stripped of the redundant degrees of freedom resulting from its being embedded in an external space, which means first of all that rigid translations and rotations should be eschewed from the picture. This attitude is reminiscent of the Leibnizian static shift argument: Taking the material content of the universe and rigidly moving or rotating it in an external Euclidean absolute space would make no empirically observable difference, so why postulating the existence of such a ``metaphysically inert'' space?

SD, however, makes a step beyond this Leibnizian attitude. Such a step is motivated by the observation that every measurement simply amounts to the comparison of two physical systems. Formally, this means that only \emph{ratios} of physical quantities carry objective information. Going back to the case of space and time, this comparativist attitude implies that the relationalist should let go also of any notion of size. What remains is simply the \emph{shape} of the system.

Mathematically, the procedure for systematically getting rid of the redundant structure associated with some symmetry is commonly known as \emph{quotienting out}. Schematically, if $\mathcal{Q}$ is the relevant configuration space of a given system and $\mathcal{G}$ is the symmetry group, the quotienting out yielding the space carrying truly physical information is $\mathcal S:=\mathcal{Q}/\mathcal{G}$, whose dimension is simply $\mathrm{dim} (\mathcal S)=\mathrm{dim}(\mathcal{Q})-\mathrm{dim}(\mathcal{G})$. In the case of $N$ classical particles, $\mathcal{Q}$ is standard configuration space, $\mathcal{G}=\mathsf{Sim(3)}$, the joint group of Euclidean translations $\mathsf{T}$, rotations $\mathsf{R}$ and dilatations $\mathsf{S}$ (called \emph{similarity group}), and $\mathcal S=\mathcal{Q}/\mathsf{Sim(3)}$ is referred to as the \emph{shape space} of the system.

Although dynamical geometry will not be considered in this paper, let us for completeness indicate the relevant symmetry group and the associated shape space. Let us denote $\mathsf{Riem}$ the set of Riemannian $3$-geometries and $\mathsf{Diff}(3)$ the group of spatial diffeomorphisms. Then, $\mathsf{Superspace}=\mathsf{Riem}/\mathsf{Diff}(3)$. Further, let $\mathcal{S}_c$ be conformal superspace $\mathcal{S}_c=\mathsf{Superspace}/\text{conformal transformations}$. Finally, the physical space of SD is $\mathcal{S}_V\equiv\mathsf{Superspace}/\text{VPCT}=\mathcal{S}_c\times\mathbb{R}^+$, where $\text{VPCT}$ is the group of \emph{volume-preserving} conformal transformations and $\mathbb{R}^+$ represents the spatial volume or its conjugate variable, the so-called \emph{York time}. The need to restrict the theory to the subset of conformal transformations that keep the volume constant arises from the empirically falsified predictions of a theory taking the full group of conformal transformations: Such a theory is unable to account for the expansion of the Universe (see \citealp{514} for details).

A precise mathematical articulation of tenets \ref{MachP2} and \ref{MachP3} yields what is today known as the \emph{Mach-Poincar\'e principle}:

\begin{ten}[\textbf{Mach-Poincar\'e principle -- classic version}]
\label{MachP}
Physical, i.e., relational initial configurations and their first derivatives alone should uniquely determine the dynamical evolution of a closed system.
\end{ten} 

Physically speaking, tenet \ref{MachP} merges Mach's idea that the fully relational character of the dynamics can be attained only when the whole universe is considered with Poincar\'e's insistence on the amount of initial data needed to describe dynamics. A distinction should be drawn between tenets \ref{MachP2} and \ref{MachP3} and tenet \ref{MachP}: The former establishes an ontological commitment towards spatial and temporal notions, whereas the latter refers to the least possible amount of initial data required to describe the dynamics. But we stress that tenet \ref{MachP} also takes on an ontological stance (as does tenet \ref{MachPSD}, which is introduced below). In this regard, the three tenets share the general methodological principle of SD, intended as an ontological commitment, namely the elimination of redundant structure, be it spatial and temporal notions (tenets \ref{MachP2} and \ref{MachP3}) or unnecessary initial data (tenets \ref{MachP} and \ref{MachPSD}).  

The use of ``first'' derivatives is problematic because it places a very strict restriction on the initial data formulation and makes it unavoidable the appearance of an external parameter, w.r.t. which \emph{extrinsic} derivatives are introduced, to account for the dynamics, be it the ratio of dilatational momenta in particles models or York time in dynamical geometry. These non-shape degrees of freedom seem to imply a commitment to physical structures other than the relational configuration itself and, thus, mark a conceptual tension with the Leibnizian/Machian spirit (See \citealp[][\S~2.2]{vasnarkos2} for a discussion on this).

The need to overcome this conceptual tension led to the latest refinement of the general framework of relationalism, dubbed \emph{Pure Shape Dynamics} (PSD; see \citealp{726}, for a general technical introduction to the framework). In a nutshell, the qualifier ``Pure'' means that PSD describes \emph{any} dynamical theory exclusively in terms of the intrinsic geometric properties---which are completely defined in terms of shape space structure alone---of the \emph{unparametrized} curve $\gamma _0$ traced out by the physical system in shape space $\mathcal S$. This ensures that there are no external reference structures nor clock processes necessary to describe $\gamma _0$ in $\mathcal S$. 
%In other words, the PSD project is highly innovative: It aims at rewriting the whole spectrum of theories of dynamics that characterized and will characterize the development of physics in fully relational, scale-invariant terms. For any ``standard'' physical system, be it a Newtonian $N$-particle system, a general relativistic cosmological model, a de Broglie-Bohm particle system, etc., PSD proposes the same recipe: Quotient out the relevant symmetries from the system and then fully geometrize its dynamical development. To further stress this insistence on the intrinsic geometric properties of the curve associated with a physical system, one speaks of its equation of \emph{state}, in contrast to its equation of motion. 
Accordingly, tenet \ref{MachP} has to be slightly modified to match PSD's ``intrinsically geometric'' nature:

\begin{ten}[\textbf{Mach-Poincar\'e principle -- modern version}]
\label{MachPSD}
Physical, i.e., relational initial configurations and their intrinsic derivatives alone should uniquely determine the dynamical evolution of a closed system.
\end{ten} 

The key innovation brought about by tenet \ref{MachPSD} is to consider in general \emph{higher-order} derivatives of the curve, thereby allowing us to describe the dynamics of a physical system in terms of the curve alone, without the need of any additional non-shape parameters as in standard SD. Clearly, there is a sense in which tenet \ref{MachPSD} is \emph{weaker} than tenet \ref{MachP}, for the former requires more initial data than the latter (higher-order versus first derivatives) to describe a dynamical system. This weakening is anything but a drawback from a modern relationalist standpoint since the upshot is the elimination of any non-shape degree of freedom from the dynamical description of a system---thus delivering a truly relational dynamics cast in terms of the degrees of freedom intrinsic to the system only. This intrinsic character of the framework is achieved through the notion of \emph{intrinsic} derivative, namely an operation that takes the derivative of variables w.r.t. parameters living in the very state space the variables belong to, unlike the extrinsic derivatives appearing in tenet \ref{MachP}. A paradigmatic example of intrinsic derivative, which is the one used within PSD, is the derivative w.r.t. the arc-length. Interestingly, any derivative w.r.t. an external parameter, be it Newtonian time or otherwise, can be exchanged for an intrinsic derivative (through, for instance, the arc-length parametrization condition; see discussion after equation \eqref{curve0}). The reason we care about intrinsic derivatives is because they provide the mathematical structure to account for the ontological requirement imposed by tenet \ref{MachP3}: Dynamics must be given intrinsically, from within the physical system. 

Given an already ``quotiented out'' physical system, we shall express the equation of state of the unparametrized curve $\gamma _0$ in its associated shape space $\mathcal S$ as follows:

\begin{equation}
\begin{array}{rcl}
   dq^a&=&u^a(q^a,\alpha _I^a)\,, \\
   d\alpha _I^a &=&A _I^a(q^a,\alpha_I^a)\,,
%   dk&=&K(q,\phi ,k)
   \end{array}
   \label{curve0}
\end{equation}

and demand that the right-hand side be described in terms of dimensionless and scale-invariant quantities, whose intrinsic change is obtained employing Hamilton's equations of motion. In \eqref{curve0}, $q^a$ are points in shape space, namely they represent the universal configurations of the system, $u^a$ is the unit tangent vector defined by the shape momenta $p_a$: 
\begin{equation*}
    u^a\equiv g^{ab}(q)\frac{p_b}{\sqrt{g^{cd}p_cp_d}}\,,
    \label{unittangent}
\end{equation*}

which allows us to define the direction $\phi ^A$ at $q^a$. It is through the unit tangent vector and the associated direction that the shape momenta enter Hamilton's equations, which are in turn used in the intermediary steps leading to the equation of state \eqref{curve0}. Finally, $\alpha _I^a$ is the set of any further degrees of freedom needed to fully describe the system. It is this set $\alpha _I^a$ that includes higher-order derivatives of the curve and spares us of the need of additional non-shape degrees of freedom. Among these, one parameter definitely stands out: A measure of the deviation of the curve from geodesic dynamics (cf. $\kappa$ in \S~\ref{dBBtechnical}, equation \eqref{ukappadefinition}). For consistency, the elements in $\alpha _I^a$ must exhaust the set of all possible dimensionless and scale-invariant quantities that can be formed out of the different parameters entering a given theory. 

The derivatives appearing in \eqref{curve0} are intrinsic, as anticipated above. Their explicit expression is $dq^a:= \frac{dq^a}{ds}$, where $s$ denotes the arc-length parameter of the curve in shape space, i.e., $ds = \sqrt{g_{ab}\,dq^a\,dq^b}$, with $g_{ab}$ the metric in $\mathcal{S}$. This expression, in turn, enables us to write the relation between intrinsic and extrinsic derivatives by means of the so-called arc-length parametrization condition: $\left(\frac{ds}{dt}\right)^2=g_{ab}\,\Dot{q}^a\,\Dot{q}^b$, with $\Dot{q}^a$ being the derivative w.r.t. the external, Newtonian time $t$. Remarkably enough, as shown in \citealp{726}, the equation of state should be taken as a whole and interpreted as giving the relative rates of change of the degrees of freedom of the curve in shape space. Thus, one has $\frac{dq^a/ds}{d\alpha _I^a/ds}=\frac{dq^a}{d\alpha _I^a}$, effectively rendering the dynamics explicitly unparametrized, hence justifying the form adopted by \eqref{curve0}.

Two crucial features of \eqref{curve0} have to be noted. First, by construction, a solution of this set of equations represents the entire history of relational evolution of the universe given certain initial conditions. Indeed, once such initial conditions are selected, \eqref{curve0} provides all the physical information needed to fully characterize the universe (i.e., a closed system): All there is to know about the universe is provided by this dynamical evolution. Second, \eqref{curve0} has an obvious unifying nature. In principle, the whole of relational dynamics, classical, relativistic, and quantum, boils down to the dynamical structure encoded in the above equation of state, which means that all these theories can be formulated in terms of some purely shape variables and their intrinsic derivatives of different orders within their corresponding shape space (Newtonian gravity, which is the example taken in \S~\ref{cm}, general relativity, which is not discussed here since it requires a paper on its own, and de Broglie-Bohm theory, which we shall explore in \S~\ref{dBBtechnical}). This is one of the most remarkable features of PSD, that is, its flexibility to describe an extremely wide range of relational motions: Simply speaking, the more structured the system, the more geometric degrees of freedom are needed to describe it.

%The essential difference with standard SD is that PSD exclusively relies on the intrinsic geometric properties of the dynamical curve in shape space, as given in \eqref{curve0}, whereas the original formulation of SD needs some extra ingredient ---be it the ratio of dilatational momenta or York time/spatial volume, as discussed earlier---to be defined on a dynamical curve in order to make sense of the physical evolution. 

%whereas standard SD does not share this prominent role of said curve. In particular, this insistence on intrinsic properties is best exhibited by the unparametrized %character of the curve in PSD, which, recall, guarantees that no external reference structures nor clock processes are needed to describe $\gamma _0$, the $\alpha _I^a$ alone being responsible for this job.

%As already stressed above, there is an important reason for considering PSD %an improvement of the original SD framework. In short, the dynamical evolution as described by PSD does not fundamentally rely on any notion of parametrization whatsoever and, hence, it is a genuinely intrinsic description of a physical system. Remarkably enough, PSD is capable of reproducing known physics despite its decidedly intrinsic nature (see \citealp{729}, for the $E=0$ $N$-body problem; the case of dynamical geometry will be the subject of another paper, currently in preparation). On the other hand, . Such a parameter, although representing a much weaker structure than Newtonian time, still represents something ``external'' to the system. 

Finally, it is important to point out how an arrow of time emerges from the dynamics of the curve in shape space: As originally put forward in \cite{706}, the function 
%cluster being a set of particles that stay close relative to the extension of the total system. Next, we demand that the complexity function grow when (i) the number of clusters do, and (ii) the clusters become ever more pronounced, namely when the ratio between the extension of the clusters to the total extension of the system grow. 
\begin{equation}
\mathsf{Com}(q)=-\frac{1}{m_{\mathrm{tot}}^{5/2}}\sqrt{I_{\mathrm{cm}}}\,V_N=\frac{\ell_{\mathrm{rms}}}{\ell_{\mathrm{mhl}}}\,,
    \label{complexity}
\end{equation}

where $I_{\mathrm{cm}}$ is the centre-of-mass moment of inertia, $V_N$ is Newton's potential and $\ell_{\mathrm{rms}}$ and $\ell_{\mathrm{mhl}}$ account for the greatest and least inter-particle separations, respectively, is a measure of the \emph{complexity} of the $N$-body system: It accounts for the formation of structures within the system, namely the extent to which particles are clustered. The remarkable feature of \eqref{complexity} is its attractor-driven behaviour, whereby the direction of secular growth of complexity \emph{defines} the arrow of time. In general, the recovery of standard temporal and spatial notions in this framework is carried out via the so-called \emph{ephemeris constructions}, which are equations relating said magnitudes to the geometric properties of the curve in shape space. In a sense, these constructions represent the ``inverse'' of the quotienting out procedure, in that they explain how and why the impression of there being an external spacetime container emerges from a fundamental dynamics that does not feature any spatiotemporal structure properly said.

The above applies to the universe as a whole, as per Mach's insights. What about subsystems? In the classical $N$-body case, as is well known, the dynamics features generic solutions which break up the original system into subsystems, consisting of individual particles and clusters, that become increasingly isolated in the asymptotic regime \citep{717}. Such almost isolated subsystems will develop approximately conserved charges, namely the energy $E$, linear momentum $\bf P$, and angular momentum $\bf J$. Within the dynamically formed subsystems, there are pairs of particles that may function as physical rods and clocks. These are referred to as \emph{Kepler pairs} because their asymptotic dynamics tends to elliptical Keplerian motion.

%This ends our necessarily brief account of classical relational theories. Our main concern in this contribution is to what extent these ideas carry over to the quantum realm. To this we turn next.

%In conclusion, PSD represents a natural evolution of Barbour and Bertotti's original ideas, which provides a robust formal framework for a full ``Machianization'' of physics. Of course, much work still has to be done but, as the previous discussion shows, the technical foundations of the theory are already laid down.

\subsection{Quantum Mechanics}\label{qm}

After the short discussion of modern relational theories in the classical case, we shall now turn our attention to a couple of attempts to extend the program to the quantum realm by focusing on non-relativistic dBB theory. The suggestion that the dBB approach is a natural choice for constructing a quantum version of SD goes back at least to \citet{433}. Indeed, the two frameworks share key insights into physical reality in that both accord a privileged status to ``spatial'' descriptions of closed physical systems, so that each dynamical stage can be seen as a snapshot of a universal configuration of material bodies. Moreover, given that the standard dBB theory is formulated on a (neo-)Newtonian background structure, the quotienting out required to render the theory relational closely resembles the procedure carried out in classical mechanics. In this section, we will be concerned with the two very first attempts at a de Broglie-Bohm-Barbour-Bettoti theory. As we shall see, both are incomplete and unsatisfactory on several grounds. %and will propose our model in \ref{dBBtechnical}.

The first model was put forward in \citet{461}, and exploited the equivalence between General Relativity (GR) and SD shown in \citet{528}.  Koslowski considered a simple quantum toy model whose classical analogue admits two formulations, one possessing key features of the spacetime description of GR, while the second sharing key features of SD. 
 
There are two main results of this comparison. First, although both quantum models yield the correct classical limit, the mechanism behind this is markedly different. The timeless quantum constraint equation of the spacetime description typical of GR is handled by taking one of its degrees of freedom as an internal clock, w.r.t. which the constrained wave function evolves. The shape dynamics description, instead, exhibits a time-dependent physical Hamiltonian, with this time parameter being the dilatational momentum $\tau$, whereby the shape dynamics analogue of the de Broglie-Bohm ``quantum potential'' decays with this time parameter.

Second, the shape dynamics analogue of the quantum potential leads to an exponentially large correction to the apparent scale at early times (when $|\tau|$ is small). Remarkably, this inflation of scale is not produced in the constrained spacetime formulation. This suggests that inflation could have a natural explanation as semiclassical effects of quantum shape dynamics. 

However promising these two features of the SD quantum model may be, there is a deep conceptual flaw, no wonder inherited from the general framework of SD known back then. We are referring to the need for an external parameter to account for dynamics. Closely related to this, this quantum model does not incorporate any measure of complexity to give rise to an arrow of time, \emph{contra} the understanding of physical evolution stressed in \S~\ref{cm}.

A second model of quantum shape dynamics has been put forward in \citep{530}, which discusses a dBB dynamics of the $N$-body system in shape space. This latter model certainly shares some formal similarities with the one we will propose in \S~\ref{dBBtechnical}, but we claim that such similarities are rather shallow, as shall be argued for presently.

First and foremost, these authors confine themselves to a purely geodesic dynamics in shape space and attempt to reproduce the physics of known regimes through gauge-fixings, effectively restoring the previously-eliminated absolute configuration space. This is physically untenable, for the authors of \citet{530} do \emph{not} embrace the fundamental insight of SD, namely, that shape space is thought of as the fundamental arena where all physics unfolds. According to this insight, laboratory physics is arrived at from the holistic fundamental dynamics in suitably effective regimes within shape space. These effective regimes allow us to write shape space as the Cartesian product of configuration spaces of subsystems, thereby effectively expressing the fundamental dynamics in terms of separate dynamical subsystems in each subspace. In particular, local frames of reference, defined through distant stars and ephemeris time, are local shape substructures within the global shape of the universe. This is in stark contrast with the spirit of \cite{530}, where shape space seems to be taken as a convenient mathematical structure wherein physics is simplest (whence their insistence on geodesic dynamics), only to get back standard configuration space physics through suitable liftings and gauge fixings within their fibre bundle structure. 

Moreover, the approach of \cite{530} implies that the wave function is non-normalizable precisely due to their insistence upon describing the interesting physics in configuration space, not shape space. This choice bars the possibility to take advantage of the fact that the shape space of the $N$-body system, the so-called \emph{shape sphere} (\citealp{514}, \S~13.1.7), is \emph{compact}, which guarantees the normalizability of the wave function. This is certainly desirable if one is to endorse the familiar interpretation of $|\Psi^2|$ as a probability distribution. The authors of \citet{530} argue that the non-normalizability arises from unphysical differences and, further, that it is the universal wave function that fails to be normalizable, while there being no reason why the conditional wave functions, the ones accounting for laboratory physics, should fail to be normalizable. This may be true, but the issue is clouded by their approach, whereas in ours---which primarily relies on the shape space formulation---the matter is readily seen.

Most importantly, these authors, like the model in \citet{461}, have not taken into account the attractor-driven behaviour of complexity, required for the emergence of an arrow of time and subsystems formation. This is the root of the \emph{physical} differences between the model of \citet{530} and ours: We shall develop a model exhibiting the modified geodesic dynamics typical of PSD, which by construction works within shape space. We will then show that this model captures the observable features of the Universe---like structure formation and the arrow of time---in appropriate regimes, without betraying the very tenets of SD. 
%Although their unparametrized curve in shape space and associated notion of relational time do certainly share some of the early insights of SD we favor, their model is not able to address the , which \emph{are} obtained in our approach thanks to the crucial attractor-driven behaviour of complexity in shape space. 

%The previous points lead us straight to our final remark, namely, that the quantum model presented in \cite{530} follows the principles of standard de Broglie-Bohm theory much closer than it does with the principles of Barbour-Bertotti theories. As a result, the D\"urr, Goldstein, and Zangh\`i's model exhibits a well-defined guidance equation \emph{by construction}, hence being able to reproduce the predictions of quantum mechanics just as standard, non-relational de Broglie-Bohm theory does. However, the approach falls short of providing a concrete step towards a fully relational quantum theory, in a Leibnizian/Machian spirit: It is more of a mathematical play---a ``shape space rendition'' of the standard theory--- rather than an attempt at a new approach to quantum physics. 

%To the contrary, our de Broglie-Bohm model follows the spirit of the SD approach very closely, with the intent of opening a window into new physics. This, as we have seen, leads to a number of open issues---the matter with the guidance equation being one of them. Of course, our refusal to ``play it safe'' may lead to a theoretical dead end but, as things stand, we regard any technical and conceptual obstacle as a natural consequence of taking a so far largely unexplored research route.

Thus, we claim that Koslowski's and D\"urr et al.'s models, although arguably do implement relational tenets, fall short of providing a satisfactory account of quantum phenomena because of \emph{conceptual} reasons (that they are ``simple'' models is a separate matter). We shall now turn to analyzing how PSD fares in this regard.

\section{A de Broglie-Bohm Model of Pure Shape Dynamics}

\subsection{Implementation}\label{dBBtechnical}

Given that the standard dBB theory describes the motion of a universal configuration of particles whose collective behavior is determined by the wave function associated with the system, a full specification of the dynamics requires two equations. First, the guidance equation, which implements the guidance principle constraining the physically allowed motions based on the wave function's form. Second, Schr\"odinger's equation, which describes the dynamics of the wave function itself. Hence, an appropriate dBB model of PSD is expected to exhibit the same dynamical features of the standard approach, but cast in terms of the geometric properties of unparametrized curves in shape space.

%So, the quantum challenge for PSD can be recast in the following terms. PSD's focus on the geometric properties of a curve in shape space can very well be thought of as an intrinsic rendering of the guidance equation: We may, in principle, describe the change in configurations employing a (possibly infinite) set of degrees of freedom. However, the guidance equation depends on an element---the wave function---that is itself subject to dynamics, which means that also this latter element should be subjected to the quotienting out procedure that strips all ``absolute'' degrees of freedom associated with a background space out of the description. Fortunately, as we shall see in a moment, the shape space of a de Broglie-Bohm $N$-body system permits to define this ``stripped down'' counterpart of the standard pilot wave. 

To obtain this, let us first define the relevant shape space, which is simply the same as in the classical $N$-body system: $\mathcal S\equiv Q^N/\mathsf T\mathsf R\mathsf S$, with $Q^N$ the configuration space of the standard dBB theory. Obviously, also the wave function must be defined on $\mathcal S$. Physically, this simply means that all information encoded in $\Psi$ regarding the embedding of the $N$-body system in an external space must be dropped. Mathematically, this is readily achieved as follows: If $\Psi _N$ is the wave function living in $Q^N$, its restriction to shape space is $\Psi _{\mathcal{S}}\equiv\,\Psi _N|_{\mathsf {Sim}}$ (strictly speaking, this reduction works only if the original wave function $\Psi _N$ is single-valued along orbits of $\mathsf {Sim}$; we shall gloss over this, given it will not affect our analysis). From now on, we will omit the label $\mathcal S$ from $\Psi$ and work with wave functions on shape space.

%In the standard, non-relational theory, our $N$-body system lives in an external $3$-dimensional Euclidean space $\mathbb{R}^3$, so we must identify all the configurations related by transformations that belong to the so-called {\emph{similarity group}} $\mathsf {Sim}$, namely the joint group of rigid translations $\mathsf T$, rotations $\mathsf R$, and dilations $\mathsf S$. Thus, the $N$-body shape space is . This completely mimics the construction of the classical case. 

%In this relational setting, the equation of state of the curve representing the succession of instantaneous configurations of the system, which are pairs $(Q^a,\Psi(q))$ of shapes and wave functions on shape space, with $Q$ standing for the actual configuration of the $N$-body system and $q$ for an arbitrary configuration in shape space. As shown in \citet{747}, the equation of state of the unparametrized curve in shape space reads:

With this technical machinery in place, the calculations readily yield (see \citealp{747}, for the technical details of the derivation):

\begin{equation}\label{debohm}
 \begin{array}{rcl}
  d Q^a &=& u^a(\phi)\\
  d \phi_A &=& \frac{\partial \Phi_A}{\partial Q^a}u^a(\phi)-\frac{\partial \Phi_A}{\partial u^a} \left(\frac 1 2 g^{cd}_{,a}(Q)u_c(\phi)u_d(\phi)+\frac 1 \kappa V_{T,a}(Q)\right)\\
  d \kappa &=& -2u^a(\phi)V_{T,a}(Q) + \tilde{K}(\kappa,\gamma,Q,\phi)\\
  d R(q) &=& -\frac{1}{\sqrt{\kappa}}\left( g^{ab}(q)R_{,a}(q)S_{,b}(q)+ \frac{1}{2} R(q)\Delta S(q)\right)\\
  d S(q) &=& -\frac{1}{\sqrt{\kappa}}\left(\frac{1}{2}g^{ab}(q)S_{,a}(q)S_{,b}(q)+V_T(q)\right),
 \end{array}
\end{equation}
where $V_T(q)=V(q)-\frac{\Delta\,R(q)}{2\,R(q)}$ is the total potential and $\Psi (q) =R(q)\,e^{iS(q)}$. Also, we have used the kinematic metric $g_{ab}(q)$ on shape space to split $S^1_a :=\left.\nabla_a\,S(q)\right|_{q^a=Q^a}$ into directions $\phi_A$ and the additional degree of freedom $\kappa$, which plays a key role in PSD, namely, a measure of the deviation of the dynamics from geodesic motion:
\begin{equation}\label{ukappadefinition}
 \begin{array}{rcl}
  u^a(\phi)&=&\frac{S^1_a}{\sqrt{g^{ab}(Q)S^1_aS^1_b}}\\
  \kappa &=&\frac{g^{ab}(Q)S^1_aS^1_b}{R^\alpha}\,.
 \end{array}
\end{equation}
In the above equation, $u^a(\phi)$ is a unit tangent vector (w.r.t. the kinematic metric $g_{ab}$) at $Q^a$ that is determined by the direction $\phi_A$, and $R$ is the scale variable appearing in a potential homogeneous in $R$ of degree $\gamma$, which is required to match the dynamics of the classical $N$-body system (see \citealp{726}). Finally, the function $\tilde{K}$ is also specified when reproducing the correct classical limit. 

To sum up, \eqref{debohm} represents the equation of state of a curve obeying the relational counterpart of the dBB laws. The curve gives the succession of instantaneous configurations of the system, which are pairs $(Q^a,\Psi(q))$ of shapes and wave functions on shape space, with $Q$ standing for the actual configuration of the $N$-body system and $q$ for an arbitrary configuration in shape space.

Note that the second term in the equation for $\kappa$ in \eqref{debohm} violates the equality between the shape momenta $p_a=\sqrt{\kappa}u_a$ and $S_{,a}$ under evolution. This violation represents a substantial departure from the standard theory, but it is not, by itself, a fatal flaw of the relational model: It just means that it is not trivial to find the regime in which the guidance principle holds, i.e., $p_a=S_{,a}$ is recovered. The physical significance of such a departure from the standard theory is easy to grasp: According to \eqref{debohm}, the emergence of subsystems satisfying Born statistics, also referred to as ``quantum equilibrium condition'', is not always possible---the guidance principle is key to ensuring it, in both dynamical relaxation and typicality arguments; see below. The investigation of which conditions  may restore Born statistics in this model is still work in progress. In other words, we are not postulating the quantum equilibrium condition, but trying to derive it from a more general dynamics. In this respect, some tentative arguments for the recovery of the guidance equation can already be given.\footnote{The gist of these arguments is due to joint work with Pooya Farokhi and Tim Koslowski.}

%Finding such a regime is important, since only in this case can a story similar to that of the standard theory be provided regarding the emergence of subsystems satisfying Born statistics (see, e.g., \citealp[][\S11.3]{235}, for a technical discussion of how to define subsystems of the universe in standard de Broglie-Bohm theory). 

%We may give some tentative arguments for the recovery of the guidance equation, which are under development with our collaborators Pooya Farokhi and Tim Koslowski\footnote{A generalisation of the model presented here, with the guidance equation \emph{and} the correct classical limit holding, is also work in progress.} 

To this extent, let us consider two conditions on subsystems: (i) Isolation, whereby the full configuration space may be written as 
$\mathcal{S}=\mathcal{S}_I\times\mathcal{S}_J\times\cdots$ for subsystems $I,J,\ldots$, and (ii) boundedness, that is, the scale $R_I$ of the subsystem $I$ tends to a constant value. From the classical case, $K_I=-\gamma\,\kappa\,\frac{D_I}{p_I}$, with $\gamma$ the homogeneity degree of the potential (see below), and $dR_I=R_I\,\frac{D_I}{p_I}$, which yield the desired result $K_I\approx\,0$. Thus, in the regime where (i) and (ii) hold, it can readily be shown from \eqref{debohm} that the guidance principle $p_a=S_{,a}$ is recovered and we may invoke either some sort of typicality considerations \citep{222} or an analog of the subquantum $H$-theorem \citep{252,253} to dynamically arrive at an effective dynamics for subsystems obeying the Born rule $\rho _I=|\psi _I|^2$, for the probability distribution $\rho _I$ and wave function $\psi _I$. Both approaches are equally viable so, as the theory stands right now, there is no indication of which of the two views should be regarded as the better one.

There is however a potential mathematical problem. Clearly, the equation of state couples $Q^a$ with $\psi$ through the quantum potential, so generally, entangled solutions to Schr\"odinger equation are expected, meaning we will have to find regimes for the existence of disjoint subsystems, $\mathcal{H}=\mathcal{H}_1\otimes\mathcal{H}_2\otimes\cdots$. This is arguably a challenging problem, whose resolution is necessary to provide a sound and robust mechanism for the emergence of subsystems. This is being actively investigated. 

Remarkably enough, the dynamics exhibited by the equation of state \eqref{debohm} shows the attractor-driven behaviour in shape space already stressed in the classical case (\citealp{706}, \citealp[][\S\S~3.5-3.6]{726}). In a nutshell, this means that the direction of secular growth of complexity---which measures the amount of structure or ``clustering'' inherent in a shape---\emph{defines} the arrow of time (see \citealp[][\S~3.4]{747}, for the numerical analysis of the 3-body system, which highlights how the attractor-driven behaviour in shape space is independent of Planck's constant). As already stressed earlier, the asymptotic evolution of the classical $N$-body system gives rise to the formation of stable substructures, among which Kepler pairs play a major role as physical rods and clocks. However, as already mentioned above, a full technical characterization of the quantum counterpart of these Kepler pairs, being part of the subtle question of the formation of subsystems satisfying effective Schr\"odinger equations in the quantum model, is still work in progress. 

%an open problem, not least because currently no compelling reason for the emergence of Planck's constant is known. A follow-up paper will tackle this crucial topic.

%The precise physical conditions that guarantee that the evolution of the quantum potential is negligible---with a relational counterpart of the standard decoherence mechanism being possibly involved---is still a subject of investigation. 

Finally, the analysis of the classical limit of \eqref{debohm} is conceptually as simple as in the standard dBB case. In a nutshell, whenever the evolution of the quantum potential $V_{Q}=-\frac{\Delta\,R}{2\,R}(Q)$ along the curve becomes negligible, the evolution of the curve in shape space is effectively described by classical equations of motion determined by the full potential $V_T=V+V_{Q}$. Hence, the classical regime is attained when $V_{Q}$ goes to zero. In this case, \eqref{debohm} reduces to the equation of state associated with the classical $N$-body system: 

\begin{equation}\label{new}
 \begin{array}{rcl}
  d q^a &=& u^a(\phi)\,,\\
  d \phi_A &=& \frac{\partial \Phi_A}{\partial q^a}u^a(\phi)-\frac{\partial \Phi_A}{\partial u^a} \left(\frac 1 2 g^{bc}_{,a}(q)u_b(\phi)u_c(\phi)
  +\frac 1 \kappa V_{,a}(q)\right)\,,\\
  d \kappa &=& -2u^a(\phi)V_{,a}(q)\mp\gamma\kappa\sqrt{-\left(1+\frac{2\,V(q)}{\kappa}\right)}\,,\\
 \end{array}
\end{equation}
where $\kappa := p^2/R^{\gamma}$, with $R^{\gamma}$ the scale component of the potential. 

\subsection{Discussion}\label{dBBphilosophical}
The dBB model of PSD just presented is a substantial step forward towards a de Broglie-Bohm-Barbour-Bertotti theory. For starters, it is not a ``shape space analog'' of standard dBB theory, \emph{contra} the spirit of the model by \citet{530}. Indeed, the fact that the guidance principle in the generalized PSD model is not automatically fulfilled signals that this model is not a mere relational rendition of the standard dBB dynamics. The novelty of the dynamical laws \eqref{debohm} resides exactly in the fact that a dBB-like dynamics is not assumed \emph{ab initio}, but it is reached when specific physical conditions are met. This means that the dBB model of PSD has (i) a broader physical significance than a standard dBB one, and (ii) it promises to shed light on the very nature of a dBB-like dynamics once a good understanding is reached of the physical conditions under which the guidance principle comes into effect.

What does \eqref{debohm} tell us about the nature of the wave function in a ``truly'' relational setting? The first thing that has to be noted is that the dynamics depicted in this system of equations features both geometrical and quantum degrees of freedom. The quantum part boils down to the phase and amplitude of the relational wave function. We immediately see that these two groups of degrees of freedom are coupled together, with none of them being mathematically reducible to the other. In this sense, the wave function in the dBB model of PSD is still a fundamental, irreducible piece of formalism that cannot be ``washed'' away. In other words, the dBB-dynamics in PSD takes on initial values of the form $(Q^a,\Psi(q))$ that are analogous to the non-relational case. However, the structure of \eqref{debohm} is still peculiar enough to suggest some features that narrow down the possible choices for a metaphysics of the wave function and the quantum degrees of freedom it represents.

To see the novel contribution that the PSD perspective on quantum physics brings, it is sufficient to ask the question about \emph{where} the quantum degrees of freedom live. Indeed, much of the debate regarding the nature of the wave function in a dBB setting touches upon the fundamental arena where the physical happenings take place. Realistic views of the wave function as a local field are often associated with a commitment to the existence of a higher-dimensional configuration space as the fundamental physical space. Likewise, stances that regard the wave function as some sort of law-like, information-driven element of the formalism tend to accord a privileged ontological status to $3$-dimensional space. In this context, the dBB model of PSD represents a third option on the metaphysical table.

At the beginning of the chapter, we said that the dynamical laws of PSD are formulated in shape space, and that an entire history of a universe governed by \eqref{curve0} is encoded in an unparametrized curve in this space, without the need of any information external to such a sequence. This suggests that shape space as a whole has physical meaning only insofar as it is considered as the collection of all physically possible curves according to \eqref{curve0}. Moreover, each and every relational configuration $q^a$ is not fundamentally placed in \emph{any} external space, be it $3$-dimensional or higher. Otherwise said, fundamentally, PSD is not committed to the existence of neither configuration space nor $3$-dimensional space---with shape space being a mere abstraction over all physically possible motions. The consequence for the debate regarding the nature of the wave function is immediate: The quantum degrees of freedom are neither ``external'' to the material ones (in the sense of there being a separate object encoding the ``quantumness'' of the world) nor ``internal''---in the sense of their being just a way to describe the evolution of the material structures. Instead, the quantum degrees of freedom are \emph{on a par} with the material ones.

What does ``being on a par'' mean from a Leibnizian/Machian relational perspective? In discussing how spatial relationalism is implemented in the classical case, we made it clear that being a material particle in that setting amounted to nothing over and above standing in spatial relations with the other \emph{relata} making up a shape. From this point of view, the standard talk of degrees of freedom of Newtonian material particles is relationally translated into talk about differences in the web of spatial relations making up the shapes ordered in a dynamical curve. Hence, there is a conceptual link between the notion of relational degree of freedom and the relations making up a shape. This is the sense in which we claim that material and quantum degrees of freedom are on a par: The fundamental relation gluing together the \emph{relata} in a dBB shape is not just spatial but a quantum/spatial hybrid. These two components are distinct, yet they are two sides of the same coin. We have argued at length elsewhere for a metaphysics of hybrid entanglement/spatial relations in PSD \citep{vasnarkos3}, so we refer the reader interested in knowing pros and cons of this choice to that paper. What interests us here is that the dBB model of PSD delivers a new take on the nature of the wave function in that it suggests that this mathematical object is ``diluted'' inside a shape.

The metaphysical moral that \eqref{debohm} suggests is, hence, that the wave function is neither an object nor a description of how quantum systems behave. Rather, it represents (part of) what it is for a system to be quantum. But isn't this going back to the ``orthodox'' view that everything there is to know about a quantum system is encoded in its quantum state? Granted, the dBB model of PSD puts an emphasis on quantum relations, not quantum states, but the worry remains that we are back to deal with metaphysically suspicious entities with a feeble connection with the manifest image of material objects inhabiting a $3$-dimensional space. To put the issue in more vivid terms, we may ask the question: How do we get from a picture of shapes being ordered ``one after another'', to a picture where a pointer on a display points in a certain direction as a consequence of measurement-like interaction with a quantum subsystem of the universe? 

The answer to this question is twofold. The first part concerns the way we get garden variety spatial and temporal notions out of a timeless sequence of shapes. The details of such a construction are carried out in \citet[][\S~4.2]{vasnarkos2}. The rough idea is that, in order to recover space and time from a purely relational description, it is sufficient to ``reverse'' the quotienting out procedure discussed in \S~\ref{cm} by using the ephemeris constructions. The physical meaning of such constructions relies on the notion of complexity, intended as a measure of the clustering among subsystems. As we saw, the growth pattern of complexity makes it possible to assign a ``time stamp'' to each and every shape in a dynamical curve, so that global notions of duration and directed flow of time can be achieved. At the same time, this construction leads to a local notion of rod and clock, as exemplified in the formation of Kepler pairs. As a result, we regain a structured picture of the universe as a spacetimeful ``arena'' for material happenings. Note how this construction does not display the conceptual obscurity with respect to notions such as location and duration that are instead inherent into standard quantum mechanics. This is for the reasons we already mentioned at the beginning of \S~\ref{qm}, that is, because the dBB framework privileges spatial concepts in the physical description, which makes it possible to accommodate the key notion of Leibnizian shape from the get-go.

The second part of the answer concerns how we can get a meaningful description of quantum subsystems whose statistical behavior matches the Born's rule of standard quantum mechanics out of ``undivided'' shapes. This can easily be done in the regime where the guidance principle holds. In this case, the decomposition of the global wave function into effective wave functions associated to subsystems can be carried out in the same vein as in the standard dBB theory.

Hence, assessing the prospects of a de Broglie-Bohm-Barbour-Bertotti theory boils down to assessing whether and to what extent this twofold answer is viable. For starters, it is still not clear what notion of complexity can be defined in a dBB context. The simple definition \eqref{complexity} works well in classical mechanics, but it should be amended to account for the more complex motions exhibited in a dBB context. This is by no means an unsurmountable problem since nothing speaks against the possibility to define a dBB-complexity that serves the same purposes of its classical counterpart in order to implement the ephemeris constructions. Instead, a more delicate issue concerns finding a recipe to define effective wave functions obeying the Born rule. First, at this stage, it is not clear under which physical conditions \eqref{debohm} satisfies the guidance principle, and which subset of such physical conditions allow for a rewriting of the wave function in terms of effective terms with mutually disjoint supports. Furthermore, it is still an open question whether the conditions under which the guidance principle is fulfilled will naturally select a probability distribution $\rho=\lvert\Psi\rvert^2$ as a typicality measure or, rather, will point toward a generic starting distribution that quickly evolves towards $\lvert\Psi\rvert^2$ as some sort of relaxation limit.

In conclusion, the model \eqref{debohm} is still far from being fully worked out but it represents a window into what is to come as far as the dBB theory is concerned. As the previous discussion hopefully highlights, by marrying the original approach to quantum physics of de Broglie and Bohm with the deep insights into relational physics of Barbour and Bertotti, it is possible to substantially progress the understanding of the quantum world, which goes far beyond the picture originally suggested in Bohm's 1952 seminal papers in terms of material particles moving under the influence of a pilot wave. The project may of course crash and burn at any point, but this is a virtue rather than a vice: It testifies that the work on quantum PSD is a legitimate physical endeavor that carries a deep ontological import, and not just an idle exercise in \emph{a priori} metaphysics. From this point of view, the prospects of a de Broglie-Bohm-Barbour-Bertotti theory look as promising and exciting as ever.

%In other words, the objection may be raised that the PSD approach to dBB theory suffers from the so-called ``problem of local beables.'' 

\section{Conclusion: On the ``Geometrizing Away'' of the Wave Function}\label{concl}

One of the key questions that motivated the research into a fully Leibnizian/Machian formulation of the dBB theory concerned the possibility to reduce the wave function to some geometric features of the framework. The essential intuition in this respect was to show that the ``influencing'' role of the wave function on the material degrees of freedom of the theory could be rendered as some geometric constraints on the physical curves in shape space. The original suggestion made in the literature \citep{433,468} was to incorporate Bohm's quantum potential into shape space's metric in a vein similar to how the gravitational potential is dispensed with in general relativity (a similar idea in the quantum gravity context dates back to \citealp{527}). The expectation was that the geodesics motions in shape space would exhibit the quantum behavior typical of the dBB theory. This target slightly changed with the onset of PSD, where the interesting dynamical curves are no longer geodesics (recall the crucial role of the ``curvature'' $\kappa$). In this new scenario, the geometrized wave function should have been rendered as some additional geometric degree of freedom of the curve besides the direction $\phi^A$ and $\kappa$. This option, however, is cumbersome to implement and, in the end, isn't really needed in order to make good sense of the model \eqref{debohm}. Does it mean that the dream of geometrizing away the wave function has been finally abandoned? Well, not really. Indeed, preliminary research shows that this is the way to go if we want to push the parsimony-driven reductionist spirit of PSD to its extreme consequences.

To better understand the direction we are heading to, recall the crucial role that the quotienting out procedure plays in PSD. This procedure permits to rigorously eliminate from the mathematical formalism any theoretical feature that is deemed redundant or metaphysically suspicious under the Leibnizian/Machian lenses. The most vivid example is of course the elimination of the degrees of freedom related to the embedding of a universal material configuration in an external space, as represented by the quotienting out of translations, rotations, and dilations. Note that this reasoning can be pushed beyond the spatial side, for example by quotienting out permutation transformations to signify that material particles bear no intrinsic identity. The metaphysical significance of the quotienting out procedure is, hence, clear: A quotiented out formal feature is something that does not belong to the fundamental ontology of the theory, although it can be recovered in the appropriate limit via an ephemeris construction.

Compare this with what happens to the standard universal wave function of the dBB theory in \eqref{debohm}. In this case, this mathematical object is not strictly speaking quotiented out; rather, it is mathematically restricted to the quotiented out space constituting the dynamical arena of the theory. This restriction is sufficient to guarantee that all and only the relational physical information encoded in the original wave function is translated into the PSD model. However, if we consider \eqref{debohm} from this perspective, it is no surprise that there is no straightforward way to reduce it to geometrical degrees of freedom of a dynamical curve: The quantum degrees of freedom inherent into the wave function are rendered relational but they are still ``out there,'' as opposed to the spatiotemporal degrees of freedom that are entirely eschewed from the fundamental ontology via the quotienting out operation.

Here lies a possible key insight: Why not quotienting out the wave function itself, instead of just restricting it to shape space? This obviously begs the question as to \emph{what} should be quotiented out in the wave function. Undoubtedly, the wave function contains a tremendous amount of information, not least the statistical information about possible measurement outcomes. So a more useful question to ask may be: What is it that \emph{shouldn't} be quotiented out? There is no definite answer to this question as yet but, intuitively, the most fundamental and indispensable information that we can think of is that encoded in the guidance principle. This is because this principle encodes the fundamental dynamical behavior of material structures. Everything over and above this---e.g., statistical information about measurement outcomes---may be seen just as a useful description to tie the fundamental behavior of matter to experimental procedures that make this behavior empirically observable.

If it should be possible to quotient out the statistical information that leads to Born's rule---and if we agree that the manifest ``quantumness'' of the world is encoded in $\lvert\Psi\rvert^2$---this would mean that the fundamental theory would be pre-quantum in a clear physical sense: It would be a theory that recovers Born's rule in an appropriate limit, thus implying that it is a non-fundamental feature of reality, but something that ``emerges'' at a certain point. If, besides this, a complete quotienting out of the spatiotemporal degrees of freedom should be possible, the end result theory would in addition be non-spatiotemporal.

Such a speculative theory would be a game changer in the quest for a theory of quantum gravity. Indeed, the common motivation for the pursuit of the many quantum gravity programs active nowadays is superseding general relativity---and, possibly, quantum field theory---by providing a quantum treatment of gravitational phenomena. The hidden premise behind this goal is that quantum physics is a window into the fundamental nature of the world. If pursued until its most extreme consequences, the PSD program would challenge this received view about quantum physics by highlighting how going quantum does not mean reaching the most basic layer of reality---in fact, quantum phenomena would not be fundamental at all. This ``final'' PSD model would constitute a radically novel framework for fundamental physics, according to which quantum and gravitational phenomena emerge from a genuinely fundamental domain that is inherently pre-quantum and does not feature the general relativistic spacetime. This is, of course, pure speculation at the moment. However, it is thrilling to see how far the original ideas of de Broglie, Bohm, Barbour, and Bertotti may lead in the quest for the understanding of the deepest nature of reality.

\begin{center}
\textbf{Acknowledgements}
\end{center}
We thank an anonymous reviewer and the editor of this volume for the very helpful comments on an earlier version of this manuscript. We also gratefully acknowledge financial support from the Polish National Science Centre, grant no. 2019/33/B/HS1/01772.

\bibliography{biblio}

\end{document}